# Auxetics-Inspired Tunable Metamaterials for Magnetic Resonance Imaging

*Ke Wu, Xiaoguang Zhao, Stephan W. Anderson *, and Xin Zhang *

K. Wu, Dr. X. Zhao, Prof. X. Zhang
Department of Mechanical Engineering, Boston University, Boston, MA 02215, United States.
E-mail: xinz@bu.edu

Dr. X. Zhao, Dr. S. W. Anderson
Department of Radiology, Boston University Medical Campus, Boston, MA, 02118, United States
E-mail: sande@bu.edu

Abstract:
**Auxetics refers to structures or materials with a negative Poisson's ratio, thereby capable of exhibiting counter-intuitive behaviors. Herein, auxetic structures are exploited to design mechanically tunable metamaterials in both planar and hemispherical configurations operating at megahertz (MHz) frequencies, optimized for their application to magnetic resonance imaging (MRI). Specially, the reported tunable metamaterials are composed of arrays of inter-jointed unit cells featuring metallic helices, enabling auxetic patterns with a negative Poisson's ratio. The deployable deformation of the metamaterials yields an added degree of freedom with respect to frequency tunability through the resultant modification of the electromagnetic interactions between unit cells. The metamaterials are fabricated using 3D printing technology and a ~20 MHz frequency shift of the resonance mode is enabled during deformation. Experimental validation is performed in a clinical (3.0 Tesla) MRI, demonstrating that the metamaterials enable a marked boost in radiofrequency (RF) field strength under resonance matched conditions, ultimately yielding a dramatic increase in the signal-to-noise ratio (SNR) (~ 4.5X) of MRI. The tunable metamaterials presented herein offer a novel pathway towards the practical utilization of metamaterials in MRI, as well as a range of other emerging applications.**

Metamaterials, defined as artificially constructed materials composed of subwavelength meta-atoms, have emerged as a promising tool to manipulate electromagnetic (EM) waves due to their extraordinary EM properties. Efforts in developing metamaterials have progressed from initial demonstrations of breaking the generalized limitations of refraction and reflection of natural materials, to their current use in facilitating a range of practical applications, such as super lenses,[1,2] perfect absorbers,[3,4] cloaking devices,[5,6] and metamaterial antennas,[7] among others. In terms of the application of metamaterials in magnetic resonance imaging (MRI), magnetic metamaterials composed of an array of unit cells featuring metallic wires or helices have been utilized to enhance the signal-to-noise ratio (SNR) of MRI by amplifying the RF magnetic field strength.[8,9] In order to achieve optimal performance, a frequency matched condition between the metamaterials and the MRI systems must be ensured. However, as metamaterials are susceptible to differences in patient body composition, their resonance frequency may shift to undesired higher or lower values when in proximity to patients of varying body composition. Another notable limitation of many conventional metamaterials reported to date for enhancing SNR in MRI systems is their planar distribution of unit cells. Though planar metamaterials provide optimal geometric coverage for a range of common imaging applications, such as spine imaging, the two-dimensional approach is limited in imaging curved surfaces such as the brain, breast, or musculoskeletal system (knee, ankle, etc.). As the SNR gains of metamaterials decrease as a function of distance from the metamaterial surface, ensuring a conformal approximation between the metamaterial and the surface of the object of interest is crucial to optimizing performance. Herein, we report the development of tunable, including both two-dimensional (2D) planar and three-dimensional (3D), hemispherical metamaterials, enabling continuous tuning of their EM properties, thereby ensuring resonance matching, in combination with the option for conformal geometries to optimize performance in MRI for a range of clinical scenarios.

Various tuning mechanisms have been both theoretically and experimentally investigated to dynamically modulate metamaterial properties, including the use of material inclusions or by changing the interactions between neighboring unit cells. Metamaterials incorporating active materials or those featuring variable impedance within or as part of the metamaterial elements have been reported to exhibit tunable EM properties, which are modulated by external influences or signals, such as optical excitation,[10] thermal heating,[11] or biased voltage.[12] Physical perturbation of the structure, by moving subsets of the unit cells fabricated on stretchable substrates[13] or using micro-electromechanical systems (MEMS) devices to adjust the orientation or relative location of a fraction of the unit cells,[14] has also been employed to yield the capacity for EM tunability. However, with respect to metamaterials for MRI applications, the focus herein, operating an additional optical or thermal source to modulate the resonance frequency is challenging given the unique environment of MRI. Furthermore, while it is possible to achieve an EM tunability by inserting variable capacitors or varactors, the requisite additional electrical circuitry required to control this system would yield an appreciable increase in complexity and cost. In the case of magnetic metamaterials, one of the most straightforward approaches to realize tunability is to adjust the separation distance between unit cells. The effect of inter-unit cell distance on near field coupling has been both theoretically and experimentally investigated, from microwave to terahertz frequencies.[15,16] However, manual geometric tuning by adjusting the inter-unit cell distance is precluded given the centimeter-scale of the unit cells and the millimeter-scale of the desired uniform physical perturbations required for frequency tuning. Therefore, we developed a novel method to realize tunability by employing a type of mechanical metamaterial, namely, auxetic cellular structures.

Auxetics refers to 2D or 3D materials and structures that preserve their global shape during expansion and contraction.[17-20] Their geometries exhibit a counter-intuitive deformation under uniaxial compression (tension) and transverse contraction (expansion), which is characterized by a negative Poisson's ratio $v$. These mechanical properties offer auxetics broad application, including advanced textiles,[21] tunable filters,[22] hierarchical stents,[23] piezoelectric devices,[24] and vibration dampers,[25] to name a few. Since the auxetic effect results from their particular internal cellular structure, structures of arbitrary geometry from 2D to 3D may be further tailored by using variable cell geometry and well-defined topologies. Furthermore, an advantage that auxetic tuning schemes possess is scalability and adaptability, properties which support the increasingly common use of these technologies. Relevant to the work presented herein, the unique properties of auxetics provide a pathway towards realizing tunable 2D and 3D metamaterials for MRI applications.

Here, we employ auxetic patterns to construct tunable 2D and 3D magnetic metamaterials composed of arrays of unit cells featuring metallic helices, capable of resonance frequency tuning for MRI. A resonance frequency tuning range (~20 MHz) is achieved using a mechanically-actuated scaffold frame to adjust the relative position between unit cells in a uniform manner, thereby modulating the coupling between neighboring unit cells. Theoretical analyses based on the coupled mode theory and finite element simulations are performed and compared to experimental results derived in a clinical (3.0 Tesla) MRI. The capacity for tuning yielded an optimized frequency match between the metamaterial and the MRI system, thereby maximizing the degree of increase in RF strength and, ultimately, the SNR of the resultant images.

With respect to the design of 2D tunable metamaterials for use in MRI, the unit cells should be arranged uniformly such that a uniform magnetic field distribution in the vicinity of the metamaterial is achieved. In addition, the constituent materials of the unit cells should be readily available in conventional 3D printing technology as this greatly eases design and process modification, thereby decreasing lead-time from concept to prototype. Considering these requirements, the rotating polygonal model type of auxetic structure is adopted in the case of the 2D tunable MRI metamaterial presented herein. The auxeticity of the rotating polygonal model results from the arrangement of rigid polygons connected together at their vertices by hinges,[26] with the deformation mechanism shown in **Figure 1**a. Upon application of an external load, the rigid polygons rotate with respect to one another, with the horizontal and vertical separation distances between neighboring unit cells simultaneously contracting or expanding and giving rise to a negative Poisson's ratio ν, which may be expressed by:[19]

$$v_{21} = (v_{21})^{-1} = \frac{a^2 sin^2\left(\frac{\theta}{2}\right) - b^2 cos^2\left(\frac{\theta}{2}\right)}{a^2 cos^2\left(\frac{\theta}{2}\right) - b^2 sin^2\left(\frac{\theta}{2}\right)} \quad (1)$$

where the geometric parameters *a*, *b* and *θ* are defined in Figure 1a. In order to yield a uniform distribution and degree of contraction or expansion in the horizontal and vertical directions for the metamaterial, here we set *a = b*, which gives rise to a negative Poisson's ratio of -1. Considering the configuration of magnetic metamaterials in MRI,[9] we employed an inner circle surrounded by four linkages to replace the polygon in the auxetic model, as shown in Figure 1a. As a proof of concept of the tunable metamaterial, cylindrical plastic scaffolds were 3D printed featuring grooves along their upper, outer surface, and used to define the shape of the helical metallic coils, including the space between turns and the number of turns. Following printing, copper wire was wound into the scaffolding grooves to form the metallic resonator composing each unit cell of the metamaterial array. In addition, at the lower, outer surface of the scaffolding, female or male supports were employed such that these unit cells could subsequently be connected by pins to form larger arrays of unit cells, with the hinge joints allowing for rotation of the unit cells, thereby giving rise to the auxetic effect; see Figure 1b. In this 2D tunable metamaterial design, the outer and inner diameters of the cylinder were 30 mm and 23 mm, respectively, and the length of male and female supports were both 3.5 mm. As a result, the separation distance *Dis* between individual unit cells could be varied within the range from 30 mm to 37 mm. Figure 1c illustrates the deformation process, demonstrating the auxetic nature of the metamaterial featuring a negative Poisson's ratio. The fabrication results may be found in Section S1 and Figure S1 in the Supporting Information.

The working principle of MRI metamaterials lies in the coupling of the helical metallic coils composing the individual unit cells. The synergy of these helical coils gives rises to a resonance mode in which the direction of the electric current induced by an applied RF field is identical in each coil. When excited by the external RF field transmitted from the MRI during RF transmission, the induced currents in the resonant mode lead to a dramatic enhancement of the RF field. When the frequency of the resonant mode closely approximates that of the working frequency of the MRI system, marked gains in both local transmit and receive RF magnetic fields are achieved, ultimately leading to enhancement in the SNR of the resultant images.[9] As above, the mechanical properties resulting from the auxetic nature of the metamaterial array allow for an adjustable separation distance d for the array. However, more directly relevant are the effects on the electromagnetic properties of the metamaterial when applied to MRI. Thus, we turn to analyses of the influence of the metamaterial's configuration on its electromagnetic properties, namely, the modulation of its resonance frequency and the variation of the magnetic field enhancement in the vicinity of the metamaterial. We consider the operating frequency (127 MHz) of the 3T clinical MRI system employed herein for subsequent experimental validation and fix the space between two turns and the number of turns of the helical coils at 1.25 mm and 7.25, respectively. Firstly, numerical simulation using CST Microwave Studio was performed, with a series of frequency responses of the metamaterial at different separation distances plotted in Figure 1d, in which resonant modes are identified as dips on the plotted curves. The resonance frequencies of this 2D tunable metamaterial were extracted from simulation results as a function of separation distance *Dis*. Moreover, theoretical analyses based on the coupled mode theory were also performed for the metamaterial. The resonant mode of the metamaterial as a function of separation distance may be derived by solving the following equation system:[27]

$$\frac{da_n(t)}{dt} = -(j\omega_0 + \Gamma_n)a_n(t) + j\sum_{k=1}^{m,k\neq n} K_{kn}a_k(t), \qquad n = 1,\cdots,m \qquad (2)$$

in which $a_k$ refers to the mode amplitude of the $k_{th}$ element in metamaterial, $\omega_0$ represents the resonant angular frequency of a single helix, $\Gamma_n$ is the intrinsic decay introduced by the material and radiation losses, $K_{kn}$ is the coupling factor among unit cells and $m$ is the total number of unit cells in the metamaterial. Details of the calculations may be found in Section S3 of the Supporting Information. Lastly, in order to validate the simulation and calculation results, we experimentally analyzed the resonance frequency as a function of metamaterial separation distance. Simulation results, theoretical calculations, and experimental results are depicted in Figure 1e, demonstrating that a ∼20 MHz frequency shift of the resonance mode may be achieved across the full range of mechanical deformation. The agreement between the theoretical calculation and experimental results serves to validate the theoretical analysis based on the coupled mode theory. The frequency match between the metamaterial and the MRI system is paramount to optimizing the enhancement of the RF magnetic field strength and, ultimately, the SNR of the resultant MRI images. The impact of frequency matching on the RF field enhancement ratio is investigated to demonstrate the necessarity of the tunability of the metamaterial. Based on simulation results, the magnetic field distribution, assuming operation at 3T MRI (127 MHz), as a function of metamaterial array deformation and resultant variations in array resonance frequency, is illustrated in Figure 1f. The leftward color map in Figure 1f depicts the magnetic field distribution on the cutting plane (indicated in the inset figure) when the separation distance $Dis$ between neighboring unit cells in metamaterial is adjusted to 30.5 mm and the corresponding resonance frequency of metamaterial is 140.8 MHz. The middle color map in Figure 1f illustrates the magnetic field distribution when $Dis$ is tuned to 33 mm, which yields an optimized resonance frequency of 127 MHz of the metamaterial array, matching the operating frequency of MRI system. The rightward color map demonstrates the magnetic field distribution when $Dis$ is increased to 36.5 mm, resulting in a metamaterial resonance frequency of 117 MHz. The dramatic weakening of the magnetic field distribution in both of the off-resonance scenarios supports the conclusion that frequency matching between the metamaterial array and the MRI system is fundamental to optimizing the field enhancement effect for the metamaterial and, ultimately, its performance in MRI. Given the clear need for frequency matching, we further conclude that metamaterial tunability is paramount to its eventual successful clinical adoption in MRI, given the known resonance shifts that may commonly be imposed on the metamaterial due to differences in patient body or phantom composition (please see Section S4 and Figure S4 in the Supporting Information for further details regarding the resonance shift due to different phantom compositions).

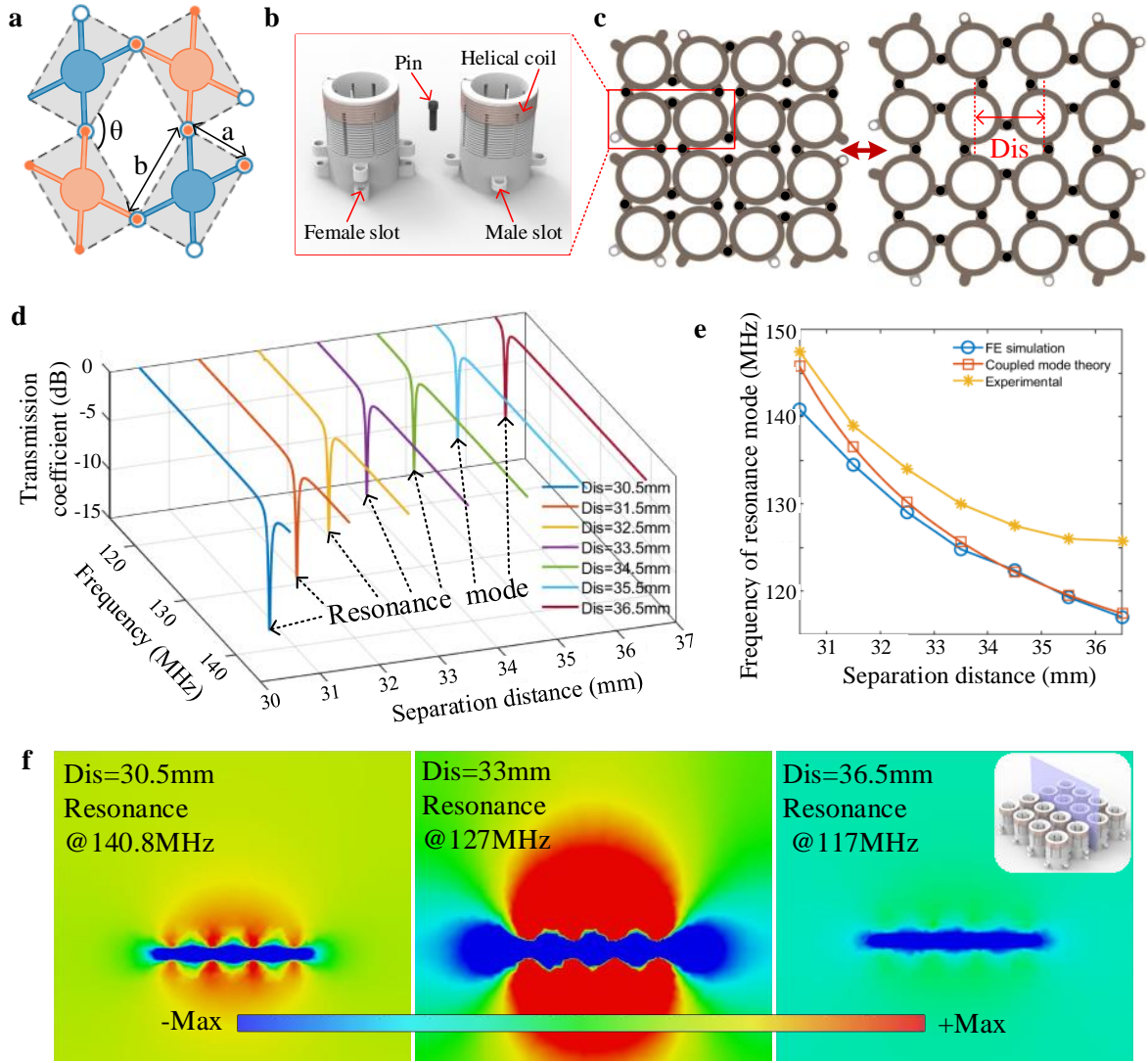

**Figure 1.** Design and electromagnetic properties analysis of the 2D tunable metamaterial. a) Schematic of 'rotating rigid polygon' type of auxetic structure. b) Illustration of structural deformation (*Dis* is defined as the separation distance between the centers of neighboring unit cells). c) Magnified schematic of the unit cells of the 2D planar tunable metamaterial. d) Transmission coefficient as a function of frequency and separation distance *Dis*. e) Tunability of the resonance mode frequency as a function of separation distance *Dis*. f) Magnetic field strength @127 MHz along the axial direction distributed along the metamaterial cross-section (depicted as the blue plane in the inset) for varied separation distances: *Dis*=30.5 mm, *Dis*=33 mm, and *Dis*=36.5 mm.

Turning our attention to 3D metamaterials, notable is the fact that 3D deployable auxetics assembled using angulated scissor elements have been developed and reported with respect to their applications in structural engineering and architecture.[28] A well-known application of 3D deployable auxetics is the Hoberman sphere,[29-31] a popular educational toy.

A Hoberman sphere is an isokinetic structure that resembles a geodesic dome, which is capable of contraction and expansion via a scissor-like action of its joints, all while maintaining its global, spherical shape. Inspired by the geometric features and deformation mechanisms of the Hoberman sphere, we developed a tunable 3D metamaterial based on the concept of an angulated scissor hinge system. More specifically, herein, we focus on the design of a hemispherical tunable metamaterial, considering its application to brain imaging and conformation to the human head. In order to distribute the unit cells of metamaterial uniformly on the surface of the hemisphere, a polyhedron tessellated with 75 isosceles triangles was modeled to form an approximately hemispherical surface, with 46 cylinders representing unit cells allocated on the vertices of these triangles. It should be noted that the lines connecting the center points of the polyhedrons $O$ to the vertices of these isosceles triangles have the same length, and the center of the polyhedron is regarded as the deployable center (modeling process is detailed in Section S2 and Figure S2 of the Supporting Information). Furthermore, the rod lengths between neighboring vertices are not perfectly equal; the rigid rods could be decomposed into three distinct types with different lengths indicated by the yellow, red and blue colors respectively, as shown in **Figure 2**a. Next, angulated scissor linkages were designed to replace these rigid rods, as shown in Figure 2b, thusly achieving the auxetic structure of the 3D tunable metamaterial. The dimensions of the angulated scissor linkages may be expressed by:

$$\beta_{1,2,3} = \pi - \frac{\alpha_{1,2,3}}{2} \tag{3}$$

$$m_{2,3} = m_1 \cdot \frac{\sin(\alpha_{2,3}/2)}{\sin(\alpha_1/2)} \tag{4}$$

$$L_{a\_1,2,3} = \frac{1}{2}[R \cdot \tan(\alpha_1/2) - m_1] \tag{5}$$

$$L_{b\_1,2,3} = \sqrt{L_{a\_1,2,3}^2 + \left(R - \frac{m_{1,2,3}}{\tan(\alpha_{1,2,3}/2)}\right)^2} \cdot \sin\left(\frac{\alpha_{1,2,3}}{2} - \arctan\left(\frac{L_{a\_1,2,3}}{R - \frac{m_{1,2,3}}{\tan(\alpha_{1,2,3}/2)}}\right)\right) \tag{6}$$

in which *α1*, *α2*, and *α3* represent the central angles formed by connecting the central point $O$ of the polyhedron to two ends of the yellow, red, and blue rods, respectively, and *β1*, *β2*, and *β3* represent the angles of the corresponding angulated scissor linkages. *m1*, *m2*, and *m3* are the offset lengths reserved for the connection between hubs and linkages. *La_1* and *Lb_1*, *La_2* and *Lb_2*, and *La_3* and *Lb_3* represent the lengths of the angulated scissor linkages replacing the yellow, red, and blue rods, respectively, as shown in Figure 2a. Details regarding design of the linkages may be found in Section S2 and Figure S3 of the Supporting

Information. Note, all relevant dimensions are illustrated in Figure 2b. In addition to the angulated scissor linkages, hubs with embedded slots were designed to serve as connection joints to support these angulated scissor linkages. Furthermore and importantly, the hubs also play the role of scaffolding for mounting the helical metallic coils of the metamaterial's unit cells. Following this geometric design process, all angulated scissor linkages and hubs were 3D printed and assembled into a deployable dome structure by connecting all the hubs through the angulated scissor linkages. Figure 2c is an illustrative example of the manner in which neighboring unit cells and their inter-connecting angulated linkages were assembled and also demonstrates the expansion and contraction process of a sample set of unit cells. Figure 2d represents a magnified illustration of the 3D tunable metamaterial, highlighting the assembly details. Figure 2e illustrates the entirety of the designed 3D deployable metamaterial, demonstrating its auxetic deformation. The fabrication results may be found in Section S1 and Figure S1 in the Supporting Information.

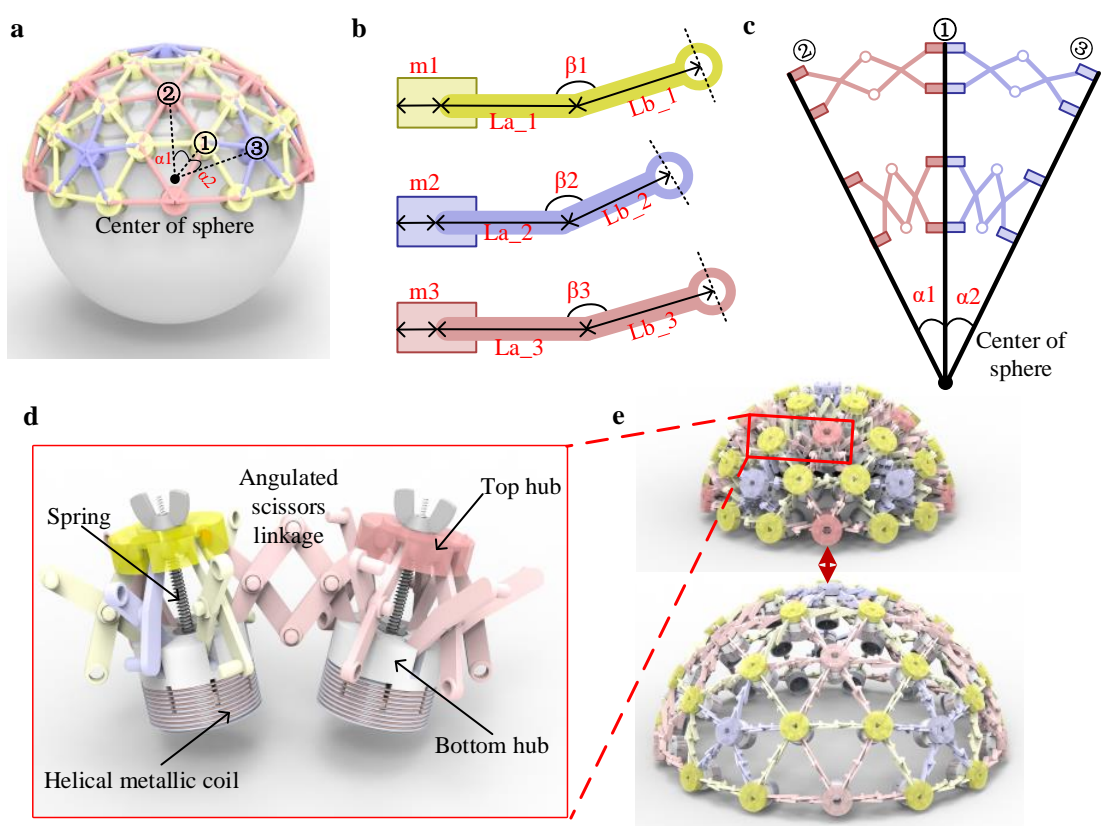

**Figure 2.** Design of the 3D tunable metamaterial. a) Modeling 3D hemispherical metamaterials. b) Angulated scissor linkages designed to replace the corresponding rods of the same color shown in (a) (dimensions indicated for subsequent analyses). c) Illustration of the manner of assembling neighboring unit cells and inter-connecting angulated linkages. d) Magnified conceptual image of discrete, inter-connected unit cells of the 3D hemispherical, tunable metamaterial. e) Illustration of the deformation of the 3D metamaterial.

Similar to the 2D tunable metamaterial, numerical simulations, theoretical analysis, and experimental validations were performed to investigate the impact of mechanical deformation on the electromagnetic properties of the 3D deployable metamaterial. Notable is that only a partial hemispherical metamaterial composed of 16 unit cells was taken into consideration when analyzing the electromagnetic properties due simply to computational limitations. The simulated frequency response as a function of the radius of the 3D metamaterial is plotted in **Figure 3**a, in which the resonant modes are identified as dips on these plotted curves. The resonance mode frequencies as a function of spherical radius *Rad* are extracted from simulation results and depicted in Figure 3b, demonstrating that a ∼20 MHz frequency shift of the resonance mode could be achieved from the auxetic deformation. For comparison, theoretical calculations based on the coupled mode theory and experimental analyses are also performed to derive the resonance frequency as a function of the radius of the metamaterial *Rad* and plotted in Figure 3b. In addition, the enhancement of the RF field strength due to the metamaterial was also simulated to investigate the impact of frequency matching between the metamaterial and the MRI system. A group of color maps of magnetic field distribution are depicted in Figure 3c, with the influence of variations in metamaterial radius and its consequent effect on metamaterial resonance frequency illustrated. As in the case of the 2D, planar version, frequency mismatch is demonstrated to strongly influence field enhancement to a degree such that the field is actually decreased in the scenario in which the metamaterials' resonance frequency is lower than the working frequency of the MRI.

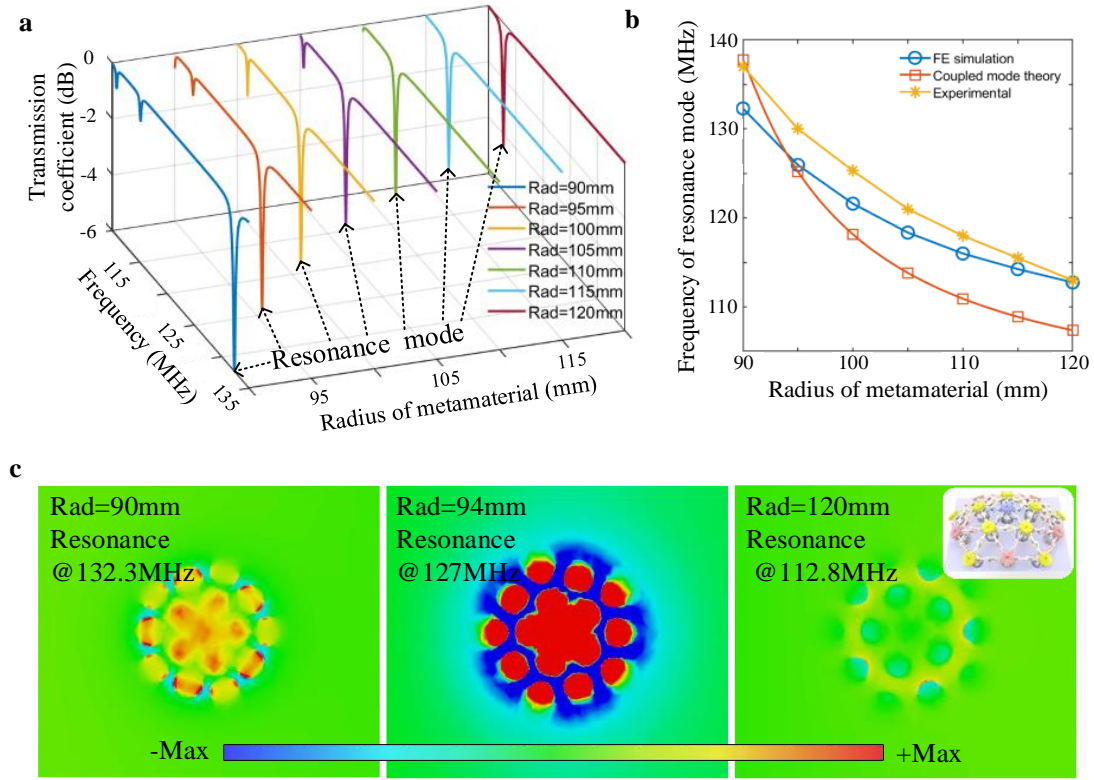

**Figure 3.** Electromagnetic properties analysis of the 3D tunable metamaterial. a) Transmission coefficient as a function of the frequency and radius of the 3D tunable metamaterial *Rad*. b) Tunability of the resonance mode frequency as a function of *Rad*. c). Magnetic field strength @127 MHz along the axial direction distributed along the metamaterial cross-section (depicted as the inset, upper right) for varied radii: *Rad* = 90 mm, *Rad* = 94 mm, and *Rad* = 120 mm, left to right, respectively.

In order to evaluate the performance of the tunable metamaterials, experimental validation of the SNR enhancement in the presence of the metamaterials was performed using phantoms in a 3T clinical MRI system. In order to calculate the SNR values for the MRI images, the two-image method was employed.[32] In the case of the 2D tunable metamaterial, a bottle-shaped phantom filled with synthetic oil was employed for imaging. The metamaterial array was placed along the undersurface of the phantom, the bottom surface of which was approximately 10 mm from the top surface of the metamaterial; the experimental setup is shown in **Figure 4**a. We initially imaged the phantom in the absence of the metamaterial, which served as a reference standard and is depicted in Figure 4b. Next, we adjusted the separation distance *Dis* in the metamaterial to its maximal value, thereby leading to a resonance frequency lower than 127 MHz. Under this frequency-mismatched state, the phantom was scanned in the presence of metamaterial, with the resultant image depicted in Figure 4c. Compared with the reference acquired in the absence of the metamaterial, there was an area of relative signal loss in the image, without any obvious improvement in the image quality. Subsequently, the resonance frequency of the 2D metamaterial was tuned to 127 MHz by decreasing the separation distance *Dis* in the metamaterial, thereby realizing the optimal condition of a resonance match between the metamaterial and the MRI system. Importantly, during the tuning process, the phantom was in proximity to the metamaterial in the identical orientation (Figure 4a) as the subsequent MRI validation in order to compensate for detuning effects related to the presence of the phantom. The image acquired under the optimized resonance state of the phantom is depicted in Figure 4d, demonstrating marked improvement in overall SNR, up to 4.5-fold when compared to the acquisition in the absence of the metamaterial. Finally, we continued to decrease the separation distance *Dis* in the metamaterial to yield a resonance frequency of the metamaterial in excess of the 127 MHz operating frequency of the MRI. The image acquired under these conditions is shown in Figure 4e, demonstrating no significant gains in SNR.

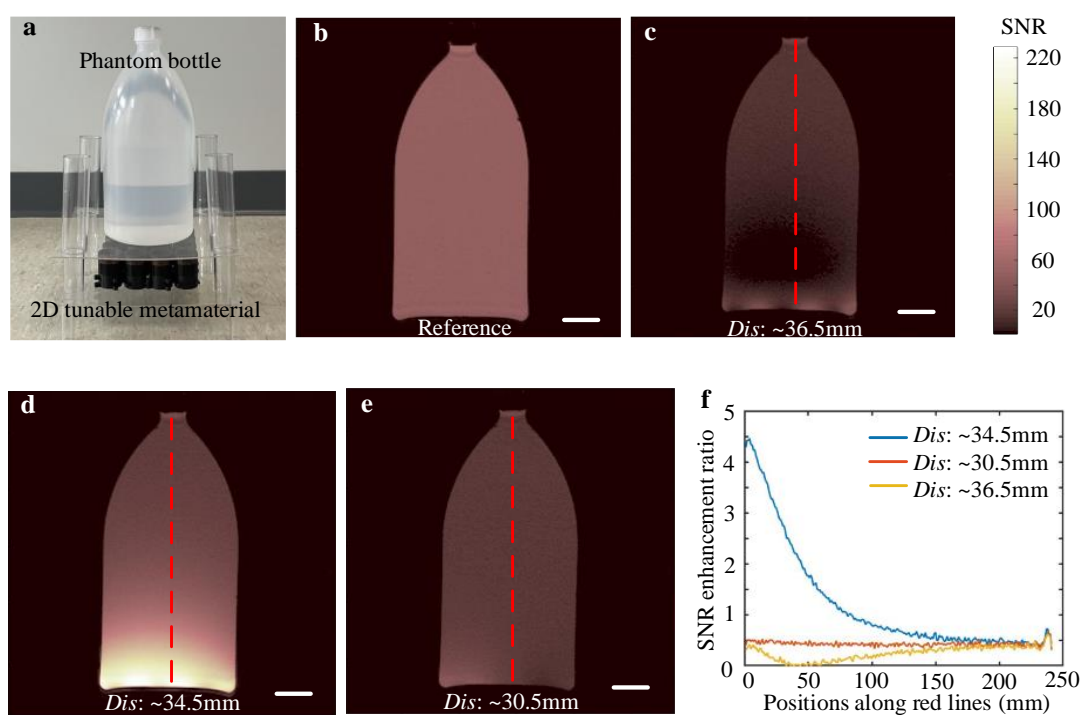

**Figure 4.** 3T MRI imaging of mineral bottle phantom. Gradient echo imaging employed. a) Photo of the experimental setup for the 2D tunable metamaterial. b) Image acquired by body coil in absence of metamaterial, serving as a reference. c) Image acquired in presence of a metamaterial with a lower resonance frequency than MRI system. d) Image captured in presence of metamaterial exhibiting a frequency match with MRI system. e) Image captured in presence of metamaterial with a higher resonance frequency than MRI system. f) Comparison of the SNR enhancement ratio for the three metamaterials of differing resonance frequency along the red dashed lines in c), d), and e). Scale bars in b), c), d), and e) are 3 cm.

Subsequently, a similar experimental approach was performed with the 3D metamaterials. In order to maintain consistency with the numerical simulation and theoretical analysis, a 3D metamaterial composed of 16 unit cells was employed in the MRI experiments. A spherical phantom filled with synthetic oil was prepared for experimental validation of the MRI SNR enhancement; the experimental setup is illustrated in **Figure 5**a. Figure 5b presents the captured image in the absence of the metamaterial, serving as a reference standard. Ranging from an expanded to a contracted state of the 3D deployable metamaterial, the resonance frequency was tuned to lower than 127 MHz, equal to 127 MHz and higher than 127 MHz, respectively, similar to the experiments performed using the 2D metamaterial version above. Under these three corresponding resonance frequencies, the spherical phantom was imaged using MRI, with the captured images depicted in Figure 5c, 5d, and 5e. As in the 2D metamaterial, the matched resonance state resulted in a maximally increased SNR in excess of 2.5-fold when compared to the reference state in the absence of the metamaterial. Also notable is the similarity in pattern between the RF field simulations in Figure 3c and the MRI images acquired using the 3D metamaterial in Figure 5d, further supporting the direct relevance between the degree of RF field enhancement and eventual SNR during image acquisition. Again, the fundamental importance of resonance matching is clearly illustrated in the case of the 3D metamaterial, as in the 2D version above, further supporting the need for the capacity for tunability in applying metamaterial technology in MRI.

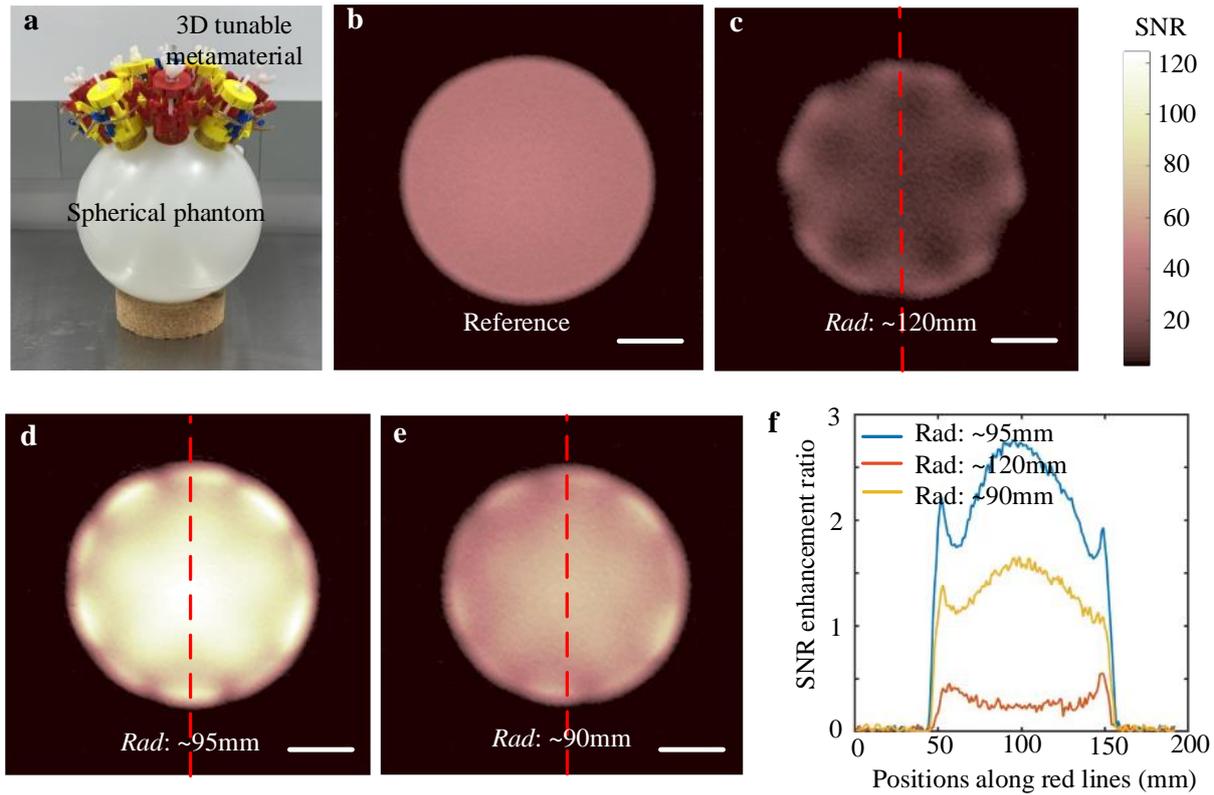

**Figure 5.** 3T MRI imaging of mineral spherical phantom. Gradient echo imaging employed. a) Experimental setup for the 3D tunable metamaterial. b) Image acquired using body coil in absence of metamaterial using spherical phantom as a reference. c) Image acquired in presence of metamaterial with a lower resonance frequency than MRI system. d) Image captured in presence of metamaterial exhibiting a frequency match with MRI system. e) Image captured in presence of metamaterial with a higher resonance frequency than MRI system. f) Comparison of the SNR enhancement ratio for the three metamaterials of differing resonance frequency along the red dashed lines in c), d), and e). Scale bars in b), c), d), and e) are 3 cm.

This paper presents the design, fabrication, and characterization of auxetics-enabled 2D planar and 3D hemispherical tunable magnetic metamaterials composed of arrays of metallic helices, which serve to enhance RF magnetic fields and increase SNR in their application to MRI. Following the introduction of the concept of tuning the resonance mode frequency by varying the density of meta-atoms through appropriately-designed 2D and 3D auxetic structures, the observed resonance shift is addressed using the coupled mode theory, simulation, and experimental analyses, deriving the influence of the metamaterials' resonance modes on their RF field enhancement effects. Lastly, MRI validations were undertaken in two clinically-relevant scenarios using planar and 3D types of auxetic metamaterials. There exist a large number of possible implementations of tunable metamaterials inspired by the concepts of auxetic structures due to their inherent design flexibility. For example, the tuning mechanism presented herein may be further developed and applied to yield arbitrarily shaped 3D tunable metamaterials, which could be employed for imaging less regular portions of the human anatomy (ankle, knee, shoulder, breasts, and neck, as examples) in a conformal fashion. Furthermore, beyond the application to MRI, auxetic behavior is theoretically scale-independent, with the same deformation mechanisms operating at the macro-, micro- and nano-levels. Therefore, this tuning mechanism offers a promising pathway for the future development of application-oriented electromagnetic devices operating across a range of designated frequencies.

**Experimental Section**

*Fabrication of the metamaterial*: In fabricating the tunable metamaterials, copper wire with a radius of 0.28 mm was wound into the scaffolding grooves to form the metallic resonator composing each unit cell of the metamaterial array. All the scaffolds were fabricated by 3D printing and subsequently assembled into 2D/3D metamaterial arrays; the fabrication results are illustrated in Figure S1.

*Characterization of metamaterial resonance frequency*: We employed a network analyzer (E5071C, Keysight Inc) with an inductive loop to excite the magnetic resonance of the 2D/3D metamaterials and measure the reflection spectra. In the reflection spectra, the dips correspond to the resonance mode of the metamaterials.

*MRI validation*: We validated the performance of the 2D and 3D metamaterials with respect to their degrees of MRI SNR improvement using a 3T MRI system (Philips Healthcare). The metamaterial was placed in the bore of the MRI system and the body coil was employed for both RF transmission and reception, as shown in Figure 4a and 5a. Gradient echo (GRE) imaging was employed with echo time (TE) and repetition time (TR) of 4.6 ms and 100 ms, respectively. During the imaging sequence of this validation, GRE imaging was first performed to capture an image of the phantom, followed by capturing a noise image by shutting down the transmission RF coil. The SNR in the region of interest was calculated as the ratio between the mean value of magnitude image and the standard deviation of the noise image, as shown in Figure S5. For the 2D metamaterial, the matrix size was 288 mm × 288 mm, the pixel size was 1 mm × 1 mm, and the slice thickness was 5 mm. A flip angle of 90° was employed in the reference (without metamaterial) imaging experiments. In the 2D metamaterial imaging experiments, the energy of the excitation RF field was reduced by a factor of 6 (flip angle = 15°) in order to ensure that the flip angle along the bottom of phantom, in proximity to the transmission RF-magnifying metamaterial, did not exceed 90°. For the 3D metamaterial, the matrix size was 192 mm × 192 mm, the pixel size was 1 mm × 1 mm, and the slice thickness was 5 mm. A flip angle of 90° was employed in the reference (without metamaterial) imaging experiments. In the 3D metamaterial imaging experiments, the energy of the excitation RF field was reduced by a factor of 3 (flip angle = 30°) in order to ensure that the flip angle along the bottom of phantom, in proximity to the transmission RF-magnifying metamaterial, did not exceed 90°.

*Numerical simulation*: The simulation of the metamaterial was performed using the finite difference time domain method with CST Microwave Studio software. In the simulation model, the dimensions of the metamaterial were the same as the fabricated sample described above. The scaffolding of the metamaterial unit cell was modeled as a dielectric material with permittivity of 2.6 and the copper wire was considered to be a metal with conductivity of $5.96 \times 10^7$ S m$^{-1}$.

**Supporting Information**
Supporting Information is available from the author.

**Acknowledgements**
This research was supported by the National Institute of Biomedical Imaging and Bioengineering grant 1R21EB024673. The authors are grateful to Dr. Yansong Zhao and Samantha Averitt for their experimental assistance during the MRI testing. The authors thank Boston University Photonics Center for technical support.

Supporting Information

**Auxetics Inspired Tunable Metamaterials for Magnetic Resonance Imaging**

*Ke Wu, Xiaoguang Zhao, Stephan W. Anderson* *, and Xin Zhang* *

K. Wu, Dr. X. Zhao, Prof. X. Zhang
Department of Mechanical Engineering, Boston University, Boston, MA 02215, United States.
E-mail: xinz@bu.edu

Dr. X. Zhao, Dr. S. W. Anderson
Department of Radiology, Boston University Medical Campus, Boston, MA, 02118, United States
E-mail: sande@bu.edu

S1. 2D/3D metamaterial fabrication

All scaffolds and linkages were individually fabricated using 3D printing. Subsequently, copper wire with a radius of 0.28 mm was wound into the scaffolding grooves to form the metallic resonators. In the case of the 2D metamaterial, 16 unit cells were assembled into a 4 × 4 array jointed by pins. In the case of the 3D metamterial, the scaffolds and linkages were printed with their constituent components in varying colors to ease the subsequent assembly process, during which 46 unit cells were assembled into a deployable, dome-shaped structure. The fabrication results are shown in **Figure S1**.

S2. Design of 3D hemispherical tunable metamaterial

In the main text, we included mathematical expressions of the dimensional parameters of the 3D metamaterial structure. Herein, we will present the geometric design process of this 3D metamaterial in a step-by-step fashion. In order to design and construct the 3D metamaterial, two fundamental considerations include ensuring both a uniform unit cell distribution, as well the capacity for a uniformly adjustable separation distance between unit cells. Based on EM simulations, the uniformity of the unit cell distribution is critical to ensuring a relatively uniform magnetic field enhancement pattern. In order to distribute the unit cells uniformly, we drew initial inspiration from the structure of C60, also known as Buckminsterfullerene, which consists of 60 carbon atoms arranged as 12 pentagons and 20 hexagons. Importantly, all of the atoms in C60 are distributed on a highly symmetric spherical surface, as shown in **Figure S2**a. We subsequently identified an additional 32 points on the spherical surface representing the intersection points between the spherical surface and straight lines passing through the sphere's center and face centers of the pentagons and hexagons. We then bisected this spherical structure to form the 3D model shown in Figure S2b. Next, all the discrete points in this structure were replaced by unit cells and connected by rods, as shown in Figure S2c. The structure in Figure S2c illustrates 46 unit cells located on the same spherical surface with a relative uniform distribution. After positioning the unit cells, the geometric dimensions of hubs were derived. Schematic illustrations of the hubs at various locations are illustrated in Figure S2d, S2e and S2f. The diameter of the hubs was set as 30 mm, with a rectangular shape ($W_s$ = 4 mm; $L_s$ = 6.5 mm) of the termini designed for subsequent of assembly of the linkages. The remaining design parameters related to the hubs are listed below.

$$\angle A_1 O_1 A_2 = \angle A_6 O_1 A_1 = 55.69° \tag{S1}$$

$$\angle A_2 O_1 A_3 = \angle A_3 O_1 A_4 = \angle A_4 O_1 A_5 = \angle A_5 O_1 A_6 = 62.15° \tag{S2}$$

$$\angle B_1 O_2 B_2 = \angle B_2 O_2 B_3 = \angle B_3 O_2 B_4 = \angle B_4 O_2 B_5 = \angle B_5 O_2 B_1 = 72° \tag{S3}$$

$$\angle C_1 O_3 C_2 = \angle C_2 O_3 C_3 = \angle C_3 O_3 C_4 = \angle C_4 O_3 C_5 = \angle C_5 O_3 C_6 = \angle C_6 O_3 C_1 = 60° \tag{S4}$$

$$O_1 A_2 = O_1 A_4 = O_1 A_6 = m_1 - 0.5 W_S \tag{S5}$$

$$O_1 A_1 = O_2 B_1 = O_2 B_2 = O_2 B_3 = O_2 B_4 = O_2 B_5 = m_2 - 0.5 W_S \tag{S6}$$

$$O_1 A_3 = O_1 A_5 = O_3 C_1 = O_3 C_2 = O_3 C_3 = O_3 C_4 = O_3 C_5 = O_3 C_6 = m_3 - 0.5 W_S \tag{S7}$$

Once the distribution of these unit cells was finalized, the angulated-scissor linkages were designed. Herein, we employ unit cells #1, #2 and #3 (shown in Figure S2c) in order to illustrate the design process for the linkages connecting the unit cells. The step-by-step construction process is detailed below:

1. Points *A*, *B*, and *C* are drawn, revealing the relative positions of the unit cells and the sphere center *O*. These points are then connected, as shown in **Figure S3**a. Even though points *O*, *A*, *B*, and *C* are not co-planar, it should be noted that drawing them on the same plane does not affect the design process for the linkages.

2. Reserve offset length *AA'*, *BB'*, *BB''*, and *CC'* for the connection between hubs and linkages, in which *AA'* = *BB'* = $m_3$ and *CC'* = *BB''* = $m_2$, as shown in Figure S3b. The relationship between $m_2$ and $m_3$ ensures that points *M* and *N* are on the same circular arc centered at point *O*.

3. Points *P* and *Q* are drawn, thus *PA'* ⊥ *MA'*, *PB'* ⊥ *MB'* and *QB''* ⊥ *NB''*, *QC'* ⊥ *NC'* respectively, as shown in Figure S3c.

4. Points *P*, *Q* and *G*, H are drawn, thus *EA'* = *FB'* = *GB''* = *HC'* = *La*. It should be noted that the value of *La* is not unique and could be any value, as shown in Figure S3d.

5. Points *E* and *F*, as well as points *G* and *H* are connected. Next, the midpoints *U* and *V*

of the line *EF* and *GH* are identified, as shown in Figure S3e.

6. After acquiring the relative positions of these points, we derived the angulated lines. By paired jointing of the angulated lines, we constructed the angulated scissor linkages, as shown in Figure S3f.

7. Figure S3g illustrates the manner by which the neighboring unit cells and interconnecting angulated linkages are assembled.

The final, assembled structure is illustrated in Figure S1, in which the deformation process is also depicted.

S3. Modeling the metamaterial frequency response

In order to derive the coupling factor $K_{kn}$ between unit cells and thereby derive the resonance frequency of the metamaterial, electromagnetic theory is initially employed to define the equivalent self-inductance and mutual inductance of the discrete unit cells as:

$$L_{ij} = \frac{\mu_0}{4\pi |I_i I_j|} \iint dr_i dr_j \frac{J(r_i)J(r_j)}{|r_i - r_j|} \tag{S8}$$

where $r_i$ and $r_j$ are the integration elements along the path of the helix, $I_i$ and $I_j$ are the equivalent electric currents in the helix, and $J(r_i)$ and $J(r_j)$ are the current densities (vectors) at $r_i$ and $r_j$. When *i* is equal to *j*, Equation S8 defines the self-inductance; when *i* and *j* differ, Equation S8 defines the mutual inductance. At the metamaterial's resonant mode, the amplitude of the electric current follows a sinusoidal profile along the helical wire, with a maximum value ($I_0$) at the center and zero at the ends of the wire; thus, $L_{i,j} = I_0/\sqrt{2}$. Both the mutual capacitance and the self-capacitance may be derived via the inverse of the coefficients of the potential matrix and be expressed as:

$$P_{ij} = \frac{1}{4\varepsilon_0 \varepsilon_r |Q_i Q_j|} \iint dr_i dr_j \frac{\rho(r_i)\rho(r_j)}{|r_i - r_j|} \tag{S9}$$

$$\mathbf{C} = \begin{bmatrix} C_{11} & \cdots & C_{1m} \\ \vdots & \ddots & \vdots \\ C_{m1} & \cdots & C_{mm} \end{bmatrix} = \begin{bmatrix} P_{11} & \cdots & P_{1m} \\ \vdots & \ddots & \vdots \\ P_{m1} & \cdots & P_{mm} \end{bmatrix}^{-1} \tag{S10}$$

in which $Q_i$ and $Q_j$ are the equivalent charge amounts in the helix and $\rho(r_i)$ and $\rho(r_j)$ are the charge densities (scalars) at $r_i$ and $r_j$. As the phase difference between the electric current and charge distribution is $\pi/2$, the charge density is at a maximum at the ends of the wire *($q_0$)* and zero at its mid-portion. With respect to the charge distribution, the charge in the helix may be expressed as $Q_{i,} = q_0 l_t \pi/\sqrt{2}$. The self-capacitance is determined by the diagonal elements of the capacitance matrix $\mathbf{C}$, while the remaining elements yield the corresponding mutual capacitance. Following derivation of capacitance and inductance, the coupling factor $K_{kn}$ and resonant angular frequency of a single helix $\omega_0$ may be readily calculated by solving the following equations:

$$\omega_0 = \frac{1}{\sqrt{L_s C_s}} \tag{S11}$$

$$K_{kn} = \frac{\omega_0}{2}\left(\frac{C_m}{C_s} + \frac{L_m}{L_s}\right) \tag{S12}$$

in which $C_m$ and $L_m$ represent the mutual capacitance and mutual inductance, respectively, and $C_s$ and $L_s$ represent the self-capacitance and self-inductance, respectively. In an effort to elucidate the factors influencing the resonant mode of the metamaterial, coupled mode theory is widely employed to describe the response of electromagnetic systems. In the case of the 2D or 3D metamaterials reported herein, given the inter-unit cell coupling, the resonant mode frequency and the resonant strength of each unit cell in the metamaterial may be derived by solving the following equation, neglecting the loss ($\Gamma_n$),:

$$\frac{da_n(t)}{dt} = -j\omega_0 a_n(t) + j \sum_{k=1}^{m, k \neq n} K_{kn} a_k(t), \quad n = 1, \cdots, m \tag{S13}$$

S4. Metamaterial resonance frequency shift due to phantoms

In order to investigate the impact of the presence of the MRI phantoms on the resonance frequency of the 2D/3D metamaterials, we employed bottle-shaped phantoms containing air, water and mineral oil (NO-TOX® Food Grade Oil, Bel-Ray) for the 2D metamaterial, and hemispherical shaped phantoms containing air, water and mineral oil for the 3D metamaterial. We employed a network analyzer (E5071C, Keysight Inc) with an inductive loop to excite the magnetic resonance of the 2D/3D metamaterials and measured the reflection spectra. In the case of 2D metamaterial, the metamaterial resonance was 127.5 MHz, 126.8 MHz and 125.5 MHz in the presence of the phantoms containing air, oil and water, respectively. In the case of 3D metamaterial, the metamaterial resonance was 127 MHz, 126.2 MHz and 123.6 MHz in the presence of the corresponding phantoms containing material air, oil and water respectively. The experimental results shown in **Figure S4** support the conclusion that resonance shifts may be imposed on the metamaterials due to differences in patient body composition. These results also demonstrate that the resonance frequency tuning ranges of our 2D/3D metamaterials are sufficient to compensate for detuning effects during imaging and, thereby, realize an optimized frequency match between the metamaterials and the MRI system.

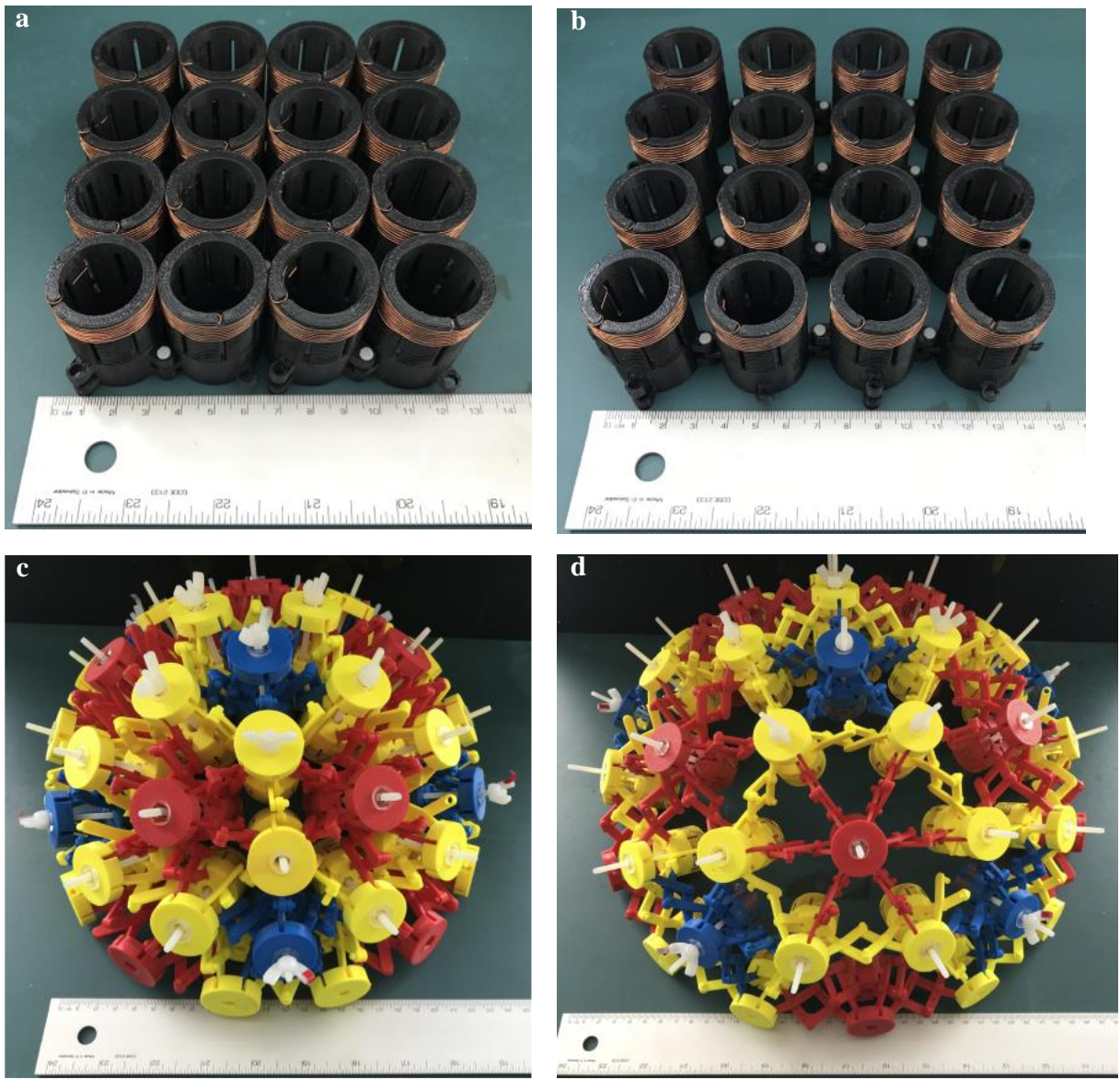

**Figure S1.** Photos of the 2D/3D metamaterials. a) Contracted state of the 2D metamaterial. b) Expanded state of the 2D metamaterial. c) Contracted state of the 3D metamaterial. d) Expanded state of the 3D metamaterial.

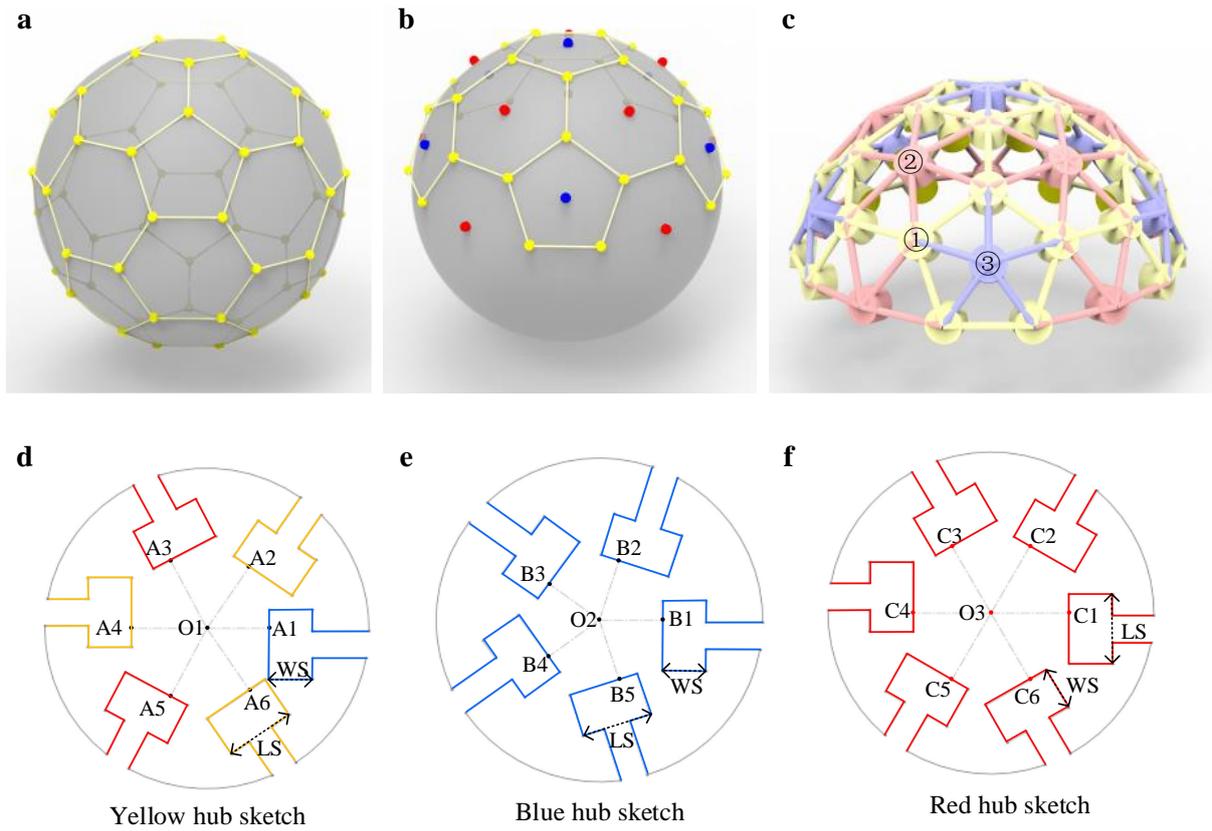

**Figure S2. Design process for structural hubs.** a) 3D spherical modeling with C60 structure. b) Interpolation of additional points for the hemispherical metamaterial. c) Distribution of unit cells of the 3D metamaterial. d) Yellow hubs' design. e) Blue hubs' design. f) Red hubs' design.

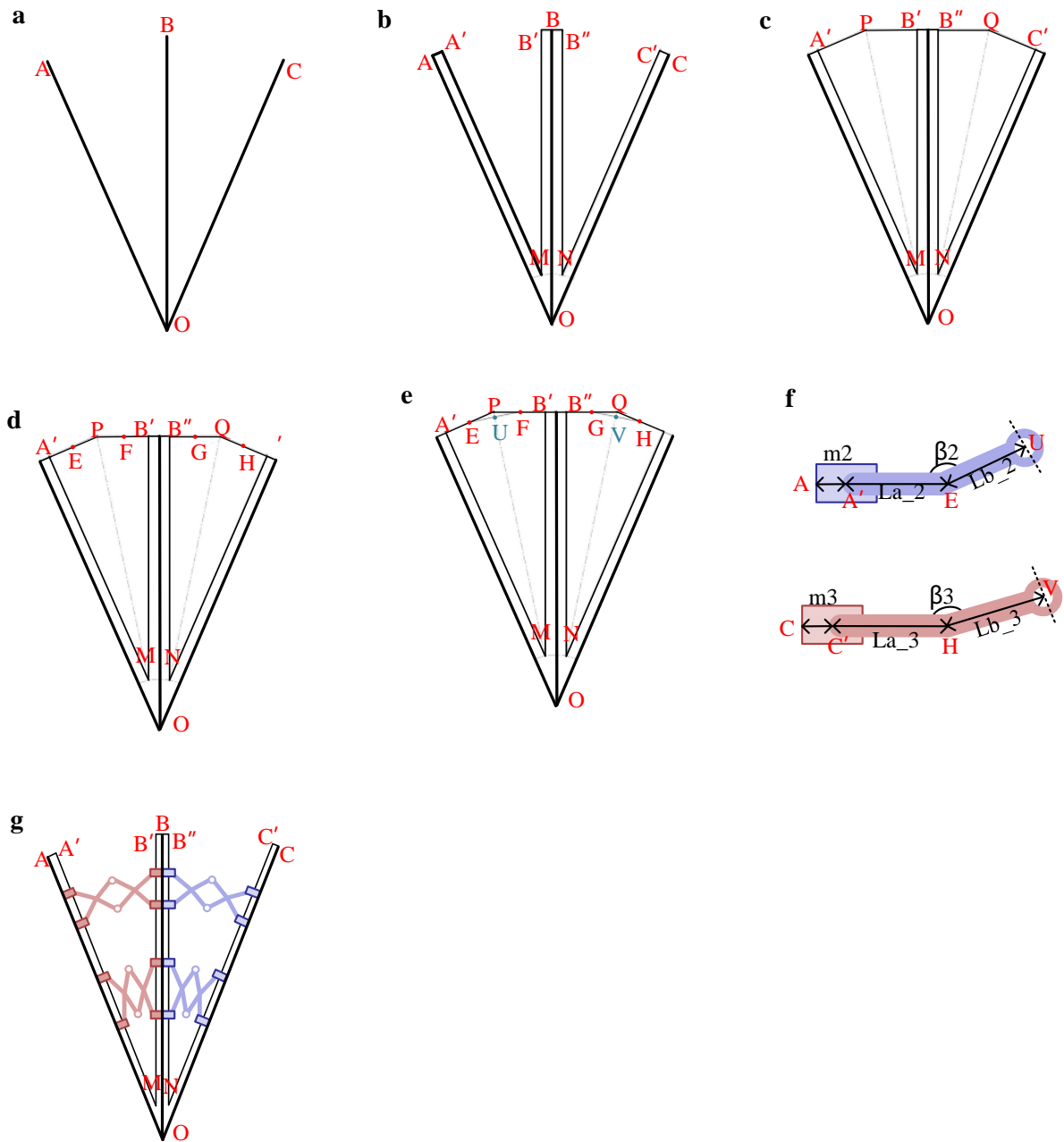

**Figure S3.** Design process for the angulated scissor linkages from a) to g). f) Design results for two distinct hubs.

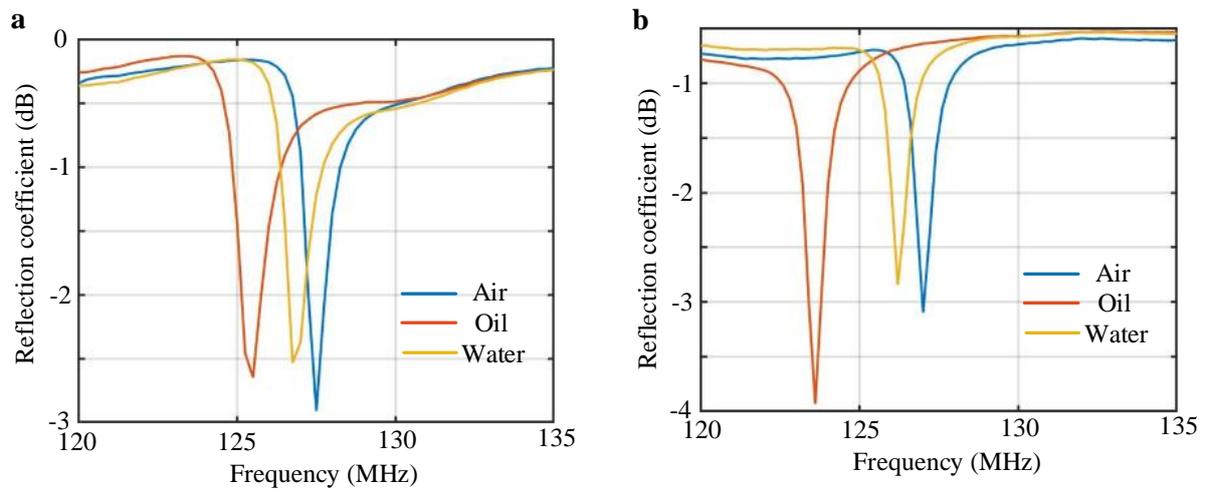

**Figure S4.** Experimental validation of the resonance shift of the metamaterials due to presence of phantoms containing different materials. a) Reflection spectra using bottle phantoms for 2D metamaterials. b) Reflection spectra tested using hemispherical phantoms for 3D metamaterials.

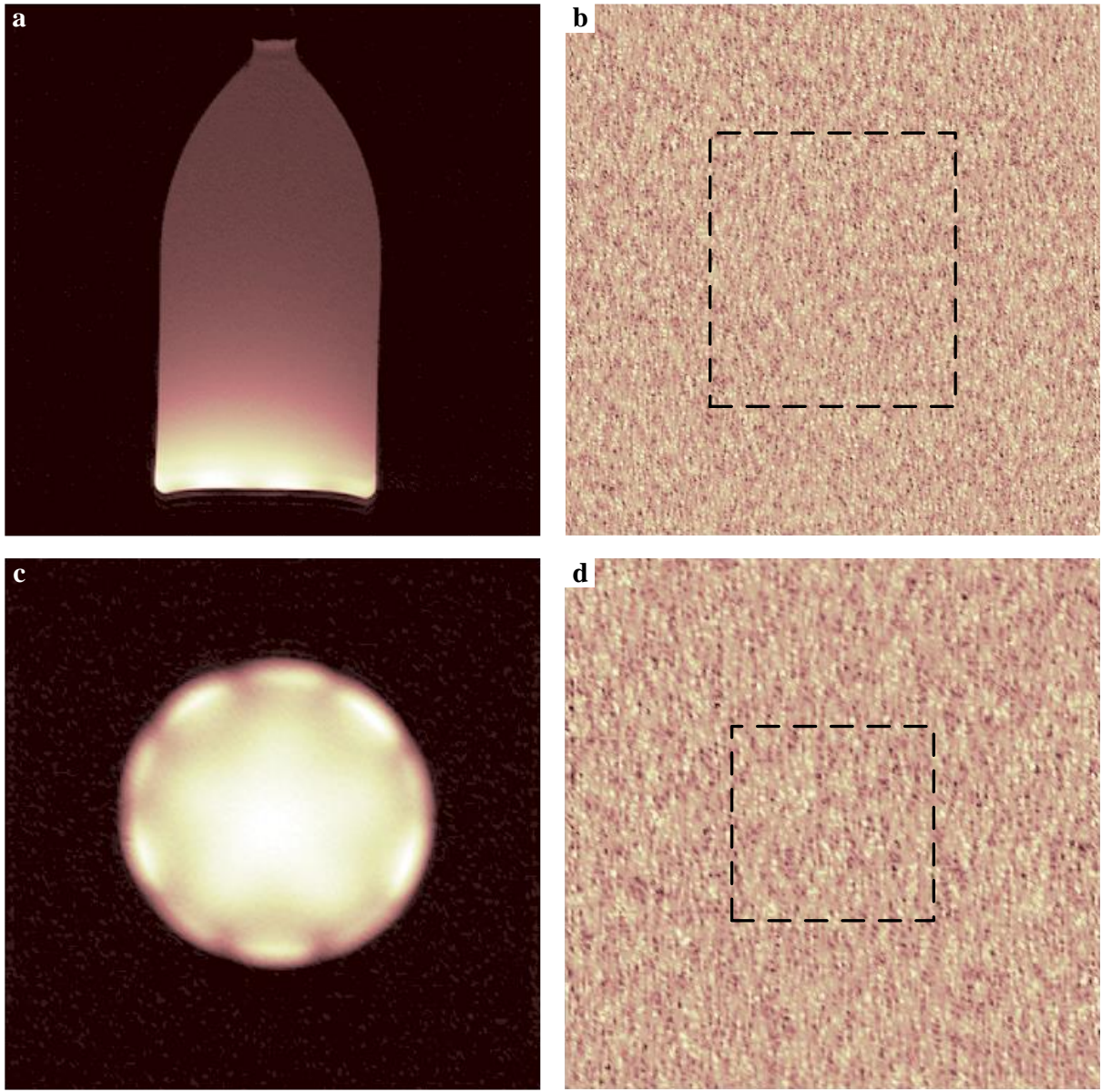

**Figure S5**. a), MRI image of the bottle-shaped phantom in the presence of the 2D metamaterial used to measure signal. b) Image captured with the transmission amplifier off, the standard deviation of the signal of which (black dashed frame) was employed to derive image noise for the 2D metamaterial. c) MRI image of the hemispherical phantom in the presence of the 3D metamaterial used to measure the signal. d) Image captured with the transmission amplifier off, the standard deviation of the signal of which (black dashed frame) was employed to derive image noise for the 3D metamaterial.